\documentclass[12pt]{article}
\begin{document}
\begin{center}
{\large{The Propagation of de Broglie Waves in Rindler Space}}

\bigskip
Soma Mitra and Somenath Chakrabarty$^\dagger$

\medskip
Department of Physics, Visva-Bharati, Santiniketan 731 235, 
West Bengal, India\\ 
$^\dagger$somenath.chakrabarty@visva-bharati.ac.in
\end{center}


\bigskip
\begin{center}
Abstract
\end{center}
In this article we have studied the propagation of matter waves in Rindler space.
We have also developed the formalism to obtained space dependent refractive index for de
Broglie waves for the particle and shown the possibility of particle
emission from the event horizon  of  classical black holes, when observed from a uniformly accelerated frame.

\bigskip
\section{Introduction}
The proposition of matter waves by Louis de Broglie \cite{R1}
during the year 1924-25 was the beginning of "new quantum
theory" or "quantum mechanics". Based on this break through hypothesis, which was later experimentally verified by
Davisson and Germar \cite{R2}, the whole quantum mechanics and
later quantum field theory were developed.
The introduction of wave functions,
uncertainty principle, quantum mechanical equations, e.g.,
Schr$\ddot{\rm{o}}$dinger equation in the non-relativistic
picture, Klein-Gordon equation and Dirac equation in the relativistic scenario for spin zero and spin half particles
respectively are all the outcome of this path breaking proposition of de Broglie. The probabilistic nature of
quantum mechanics, which is the main difference from the 
deterministic classical idea is also the contribution of de Broglie's
hypothesis of matter wave. A new era in modern physics was started
after this hypothesis. The electrons in atomic orbitals are replaced by the electron clouds, which were
particles in the original Bohr's model.

The Schr$\ddot{\rm{o}}$dinger equation may be thought of a diffusion
equation for the de Broglie waves in a complex space-time coordinate
space, or grossly speaking in Hilbert space, where we do not have
direct access. However, the basic properties of the space-time coordinates are
Euclidean in nature. Whereas the Klein-Gordon equation and Dirac
equation are in Minkowski space, i.e., in flat space-time coordinate
space of special theory of relativity.

In the present article we shall investigate the propagation of de
Broglie waves in a flat space but observed from a uniformly
accelerated frame. The space is called the Rindler space \cite{R3}.
In some very recent published articles we have reported the formalisms
to obtain the stationary state solutions for the Schr$\ddot{\rm{o}}$dinger equation in
Rindler space \cite{R4,R5} and predicted a new kind of quanta,
which is named as COSMIC PHONON by us. 

The single particle Hamiltonian in Rindler space is found to be
non-Hermitian. However, it is $PT$-symmetric \cite{R4,R5}. As a
consequence the energy eigen values or eigen spectrum are found to be real in
nature \cite{R6} (see also \cite{R4,R8} Recently we have developed an exact formalism to obtain the
solutions for the relativistic version of single particle quantum
mechanical equations in Rindler space \cite{R8}. 

In the present article we shall study the propagation of de Broglie
waves in Rindler space, which is exactly like the Minkowski space
except the observed frame is moving with uniform acceleration. The
space is also flat in nature. To the best of our knowledge this
problem has not been studied before. 

The article is organized in the following manner. In the next
section we have developed the basic formalism and in the last
section we have given the conclusion of this work.
\section{Basic Formalism}
In the Rindler space the square of the geodesic is given by
\cite{R9,R91,R10,R10a}
\begin{equation}
ds^2=\left (1+\frac{\alpha r}{c^2}\right )^2 c^2dt^2-dr^2
\end{equation}
where the three-space is assumed to be isotropic and spherically
symmetric. Here $\alpha$ is the uniform acceleration of the frame and
$c$ is the speed of light in free space. Hence from the minimization
of action, the single particle Lagrangian is given by \cite{R11}
\begin{equation}
L=-m_0c^2\left [ \left ( 1+\frac{\alpha r}{c^2} \right
)^2-\frac{v^2}{c^2} \right ]^{1/2}
\end{equation}
where $m_0$ is the rest mass and $v$ is the three
velocity of the particle.

Before we proceed further, we would like to introduce the principle of
equivalence. According to which a frame undergoing an accelerated
motion is equivalent to a rest frame in presence of a gravitational
field. The strength of this field is exactly equal to the magnitude
of the acceleration of the frame. Therefore equivalently speaking, in
the present article we have studied the effect of uniform gravitational field
on the propagation of de Broglie waves. In particular we would
like to see the effect of ultra-strong gravitational field on the
propagation of de Broglie waves very close proximity of the surface of a
stellar black hole or near the event horizon. The effect of strong gravitational field
has been emulated through the modification of refractive index of the
medium and the speed of matter waves. Analogous to the general theory
of relativity, where in presence of gravitational field the
vacuum or free space becomes a space varying refracting medium \cite{R12} for electromagnetic waves, in
the present scenario also there are spatial variation of refractive 
index of de Broglie waves and the speed of matter waves. In the
subsequent analysis of this article, we shall obtain the analytical form of
such spatial variations of refractive index and the speed of matter
waves, when observed from a uniformly accelerated frame. 

From the single particle Lagrangian (eqn.(2)), the three momentum is
given by
\begin{equation}
p= \frac{\partial L}{\partial v}= \frac{m_0 v}{\left [\left ( 1 +
\frac{\alpha r}{c^2}\right )^2 - \frac{v^2}{c^2}\right ]^{1/2}}
\end{equation}
Hence one can write down the Hamiltonian of the particle in the
following form
\begin{equation}
H=\vec p.\vec v-L = \left ( 1 + \frac{\alpha r }{c^2}\right )\left (p^2 c^2 + m_0 ^2
c^4 \right )^{1/2}
\end{equation}
Combining eqn.(2) and  eqn.(3), we have 
\begin{equation}
\frac{v}{c} = \frac{pc\left ( 1 + \frac{\alpha r }{c^2}\right
)}{\left (p^2 c^2 + m_0 ^2 c^4 \right )^{1/2}},
\end{equation}
and 
\begin{equation}
p = \frac{H}{c} \frac{\left [ 1 - \frac{m_0 ^2 c^4}{H^2}\left ( 1 +
\frac{\alpha r }{c^2}\right )^2\right ]^{1/2}}{1 + \frac{\alpha r}{c^2}}
\end{equation}
Writing $ p = \hbar k $, where $ k = \frac{2 \pi }{ \lambda } $, the 
wave-number and $ \lambda $ is the wavelength for the de Broglie wave 
associated with the particle, then we have  
\begin{equation}
\lambda (r) = \frac{hc( 1 + \frac{\alpha r}{c^2})}{H [ 1 - \frac{m_0
^2 c^4}{H^2}( 1 + \frac{\alpha r}{c^2})^2]^{1/2}}
\end{equation}
Therefore exactly like the electromagnetic case, the de Broglie wavelength  is also space dependent. 
Now for photon, the quantized form of electromagnetic wave, the
rest mass $ m_0 = 0 $, hence 
\begin{equation}
\lambda (r) = \frac{hc( 1 + \frac{\alpha r }{c^2})}{h \nu}
\end{equation}
the space dependent wave length for the electromagnetic case.
Here exactly like the photon case we put $ H = h\nu $ for the de
Broglie waves. Since the matter  wave is
propagating in a uniform gravitational field, the energy $H$ is assumed to be a
constant. Further the constancy of energy of the particle is maintained by
the space variation of wave length and speed of the de Broglie wave
associated with the moving particle.

Then it is a matter of simple algebra to show that for the
electromagnetic waves or for the photons
\begin{equation}
\nu \lambda (r) = c( 1 + \frac{\alpha r}{c^2}) = \frac{c}{n(r)}
\end{equation}
the space dependent refractive index, where
\begin{equation}
n (r) = \frac{c}{c(r)} = \frac{c}{( 1 + \frac{\alpha r }{c^2})}
=\frac{1}{1+\frac{\alpha r}{c^2}}
\end{equation}
and
\begin{equation}
c(r)=c\left ( 1+\frac{\alpha r}{c^2}\right )
\end{equation}
the space dependent speed of light. This is analogous to the space
variations of refractive index and the speed of light 
in presence of gravity as has
been discussed in general theory of relativity \cite{R12}.

Therefore we can rewrite equ.(8) as 
\begin{equation}
\lambda (r) = \frac{hc}{n(r)H \left [ 1 - \frac{m_0 ^2 c^4}{H^2}\left ( 1 +
\frac{\alpha r}{c^2}\right )^2\right ]^{1/2}}
\end{equation}
Hence one can  have
\begin{equation}
\lambda (r) n(r) \left [ 1 - \frac{m_0 ^2 c^4}{H^2}\left ( 1 + \frac{\alpha
r}{c^2}\right )^2\right ]^{1/2} = \frac{hc}{H} = \rm{constant}
\end{equation}
Further introducing the  
space dependent refractive index for de Broglie wave in the refractive medium in 
presence of gravity as
\begin{equation}
 \mu (r) = n(r) \left [ 1 - \frac{m_0 ^2 c^4}{H^2}\left ( 1 + 
 \frac{\alpha r}{c^2}\right )^2\right ]^{1/2}
\end{equation}
we have
\begin{equation}
\lambda (r) \mu (r) = \rm{constant}
\end{equation}
Hence we can see that for $ m_0 = 0 $, i.e., for photons 
\begin{equation}
\lambda (r) n(r) = \rm{constant}
\end{equation}
With eqn.(11) for $n(r)$, we have 
\begin{equation}
\mu(r)=\left [ \frac{1}{\left (1+\frac{\alpha r}{c^2}\right )^2} -
\frac{m_0^2c^4}{H^2}\right ]^{1/2}
\end{equation}
To have some meaningful variation of $\mu(r)$ with $r$, we write
$\alpha=-GM/r^2$ assuming Newtonian approximation,  
where $M$ is the mass of the strongly gravitating
object. Then the refractive index can be written as
\begin{equation}
\mu(r)=\left [ \frac{1}{\left (1-\frac{GM}{rc^2}\right )^2} -
\frac{m_0^2c^4}{H^2} \right ]^{1/2} 
\end{equation}
Here again like the electromagnetic case the refractive index of the medium
increases with the strength of gravitational field. Since $\mu\equiv \mu(r)$, the de Broglie wave also can not
follow a straight line in presence of gravity. The path will be a geodesic.
Since the refractive index changes continuously with the radial coordinate $r$, the produced particles and also
photons (with $m_0=0$) will slowly deviate from the normal 
drawn each and every points in the space out 
side the gravitating
object. The normals are along the radial direction at every point.
This is also true even if the gravitating object is black hole.
Now to see the effect of ultra strong gravitational field near the
event horizon of a black hole, we put $R_s=2GM/c^2$ as the
Schwarzschild radius in eqn.(18), where $M$ is now the mass of the
black hole. Then we have
\begin{equation} 
\mu(r)=\left [ \frac{1}{\left ( 1-\frac{R_s}{2r}\right )^2} -
\frac{m_0^2c^4}{H^2}\right
]^{1/2}
\end{equation}
Therefore at $r=R_s/2$, the first term within the third bracket
becomes infinity. The second term is obviously $<1$. This term,
$\mu(r)$,
therefore
goes to infinity as $r\longrightarrow R_s/2$.
This is also true for the photon case. 
In this limit the refractive index $\mu(r)$ 
for the electromagnetic waves and also for the 
de Broglie waves is positive infinity. But the
point $r=R_s/2$ is within the event horizon and is not accessible to
any observer who is outside the event horizon. However near the event
horizon, i.e., for $r\longrightarrow R_s$, we have for the de
Broglie wave
\begin{equation}
\mu(r)=\left [4-\frac{m_0^2c^4}{H^2}\right ]^{1/2}
=2\left [1-\frac{m_0^2c^4}{4H^2}\right ]^{1/2}
\end{equation}
which is obviously less than two. Whereas for the photons this
quantity is exactly equal to two.
Therefore considering a hypothetical situation, where the
refractive index is unity on the other side of the event horizon,
the critical angle of incidence near the event horizon from
inside and just for the grazing emergence of the same matter wave on the other side is given
by \cite{R12a}
\begin{equation}
\theta_r=\sin^{-1}\left [ \frac{1}{
2\left (1-\frac{m_0^2c^4}{4H^2}\right )^{1/2}}
\right ]
\end{equation}
which is $<30^0$.
However, it is exactly $30^0$ for electromagnetic waves with $m_0=0$.
Therefore in analogy to the emission of electromagnetic waves, 
for matter waves also there is a cone like
window of solid angle $\sim 60^0$ subtended by this spherical window
element at the point of emergence near the event horizon. Therefore
among the emitted particles from this point at various directions
with different angles, only
those can escape the event horizon whose incident angles are
within this window of solid angle $\sim 60^0$. However there are infinite
number of such points on the black hole surface at which the
particles are created. Therefore there will also be infinite number
of such windows. A large number of them are overlapping with each
other. As a consequence for an observer in a frame undergoing uniform
accelerated motion, the surface at the close proximity of event
horizon is almost transparent for the emission of matter waves associated
with the produced particles and also with the emitted photons. 
Therefore, there will be emission of particles and also photons from the
surface of  classical black holes when observed from a uniformly
accelerated frame. However, for de Broglie waves the solid angle  will be
$<60^0$, but $>0^0$. On the other hand for the photons it is exactly
$60^0$.
\section{Conclusion}
Therefor we may conclude that
analogous to Unruh effect \cite{R14a,R14b} in which the inertial
vacuums are warmer or hotter as observed from uniformly accelerated
frames, the classical black holes are no longer black, but emit
electromagnetic radiation and also particles.
However one can show very easily that in the case of Schwarzschild 
geometry, there can not be any kind of
emission from classical black holes \cite{R14}.

\end{document}